\documentclass[12pt]{iopart}
\usepackage{graphicx}
\usepackage{iopams}
\usepackage{citesort}
\usepackage{accents}
\usepackage{url}
\graphicspath{{./}{paper_data/}}

\newcommand{\OleMiss}{Department of Physics and Astronomy,
		University of Mississippi, University, Mississippi 38677, USA}
\newcommand{\NBI}{Niels Bohr International Academy, 
		Niels Bohr Institute, Blegdamsvej 17, DK-2100 Copenhagen, Denmark}
\newcommand{\WFU}{Department of Physics, 
        Wake Forest University, Winston-Salem, North Carolina 27109, USA}
\newcommand{\UB}{Institute for Gravitational Wave Astronomy \& School of Physics and Astronomy, 
        University of Birmingham, Edgbaston, Birmingham B15 2TT, UK}
\newcommand{\Caltech}{TAPIR,
		California Institute of Technology, Pasadena, CA 91125, USA}

\begin{document}

\note[]{Multimode ringdown modelling with \texttt{qnmfits} and \texttt{KerrRingdown}}
\author{L Magaña Zertuche$^{1,2}$, L Gao$^3$, E Finch$^{4,5}$, G B Cook$^3$}
\address{$^1$ \OleMiss}
\address{$^2$ \NBI}
\address{$^3$ \WFU}
\address{$^4$ \Caltech}
\address{$^5$ \UB}

\eads{\mailto{lorena.zertuche@nbi.ku.dk}, \mailto{gaol18@wfu.edu}, \mailto{efinch@caltech.edu}}
\vspace{10pt}
\begin{indented}
\item[]February 2025
\end{indented}

\begin{abstract}
    In the last decade, the ringdown community has made large strides in
    understanding the aftermath of binary black hole mergers through the study 
    of numerical simulations. In this note, we introduce two flavors of fitting
    algorithms, that have been verified against each other, for the extraction
    of quasinormal mode amplitudes from ringdown waveforms — \texttt{qnmfits} 
    in Python and \texttt{KerrRingdown} in Mathematica. 
\end{abstract}

%
\vspace{2pc}

\submitto{\CQG}
%
%
%

\section{Introduction}

The ringdown is the final stage of binary black hole (BH) coalescence, starting near the peak intensity of a gravitational wave (GW) signal. 
In the regime of linear perturbation theory, the ringdown is described by a sum of quasinormal modes (QNMs), for which the angular functions are the spin-weight $-2$ spheroidal harmonics with indices $(\ell,m,n)$~\cite{Teukolsky:1972my,Teukolsky:1973,Press:1973zz}. 
The time-dependent part of each QNM is an exponentially damped sinusoid with a characteristic complex frequency, the real part of which gives the angular frequency and the imaginary part giving one over the damping time.
For astrophysical BHs within general relativity (GR) this complex frequency is completely determined by the BH mass $M$ and spin $\chi$, giving rise to a QNM frequency spectrum seen in Fig.~\ref{fig:qnm_taxonomy}~\cite{Finch:2023}.
Each QNM also has an associated amplitude and phase, which will depend on the astrophysical process causing the perturbation.
This means (in principle) that the excitation of these QNMs encodes the properties of the binary system before the ringdown. 

The simplicity of the ringdown makes it an ideal setting for tests of GR and for learning about the properties of both the progenitor and remnant BHs.
However, there remain many open questions about the ringdown signal. 
These include the importance of overtones~\cite{Giesler:2019uxc} and mirror modes~\cite{Dhani:2020nik}, the unknown ringdown start time, QNM spectral instability~\cite{Cheung:2021bol,Berti:2022xfj,Cardoso:2024mrw}, the impact of a dynamically changing spacetime through the ringdown~\cite{Redondo-Yuste:2023ipg,Zhu:2024dyl}, nonlinear QNMs~\cite{Mitman:2022qdl,Cheung:2022rbm}, and the exact mapping from progenitor to remnant properties~\cite{London:2014cma,London:2018gaq,Cheung:2023vki,Hamilton:2023znn,Zhu:2023fnf} (particularly for systems with precession~\cite{OShaughnessy:2012iol,Pratten:2020ceb,Hamilton:2023znn,Zhu:2023fnf}). 

One method to shine light on these issues is to perform fits to the ringdown from numerical-relativity (NR) simulations. 
NR waveforms are a powerful tool for exploring what the ringdown looks like within GR, and studies of their QNM content inform our expectations when analyzing real GW data.
Several codes have been developed for this purpose, such as the agnostic fitting of Refs.~\cite{Baibhav:2023clw,Cheung:2023vki} or the QNM filter of Ref.~\cite{Ma:2022wpv}.
More recently, a Bayesian approach to fitting NR has also been employed by Refs.~\cite{Redondo-Yuste:2023seq,Carullo:2023tff,Clarke:2024lwi}.
Perhaps the most widely used method is to perform a least-squares fit to extract QNM amplitudes and phases; in this note, we present easy-to-use Python and Mathematica codes for performing multi-mode, i.e., all-sky, least-squares fits to the ringdown of NR waveforms: \texttt{qnmfits}~\cite{qnmfitscode} and \texttt{KerrRingdown}~\cite{KerrRingdownCode}.
Our goal is to provide robust ringdown-fitting packages that can be used for comparison or further investigation of ringdown modeling; these codes have been written independently and verified against each other.

Several NR catalogs are available~\cite{Boyle:2019kee,Healy:2022wdn,Hamilton:2023qkv,Ferguson:2023vta}, and our codes can be used to fit the waveforms from any of these catalogs or any other generic waveform data.
However, the code has been developed with SXS waveforms in mind (the Python code leverages the \texttt{sxs} and \texttt{scri} packages~\cite{Boyle_The_sxs_package_2025,scri_url}), and in this note, we demonstrate the codes with Cauchy-characteristic Evolution (CCE) waveforms mapped to the superrest frame~\cite{MaganaZertuche:2021syq} (known to be optimal for the ringdown).

The outline of this paper is as follows. 
Section~\ref{Theory and Conventions} presents background theory and the relevant conventions used throughout the paper. 
Section~\ref{Demonstration} presents some simple results obtained with the codes, which act as a demonstration of their capabilities.
Section~\ref{Conclusions} presents a discussion of the results described in the previous section and the conclusions.

\section{Theory and Conventions}\label{Theory and Conventions}

\subsection{QNM Frequencies}

\begin{figure}
	\centering
	\includegraphics[width=\columnwidth]{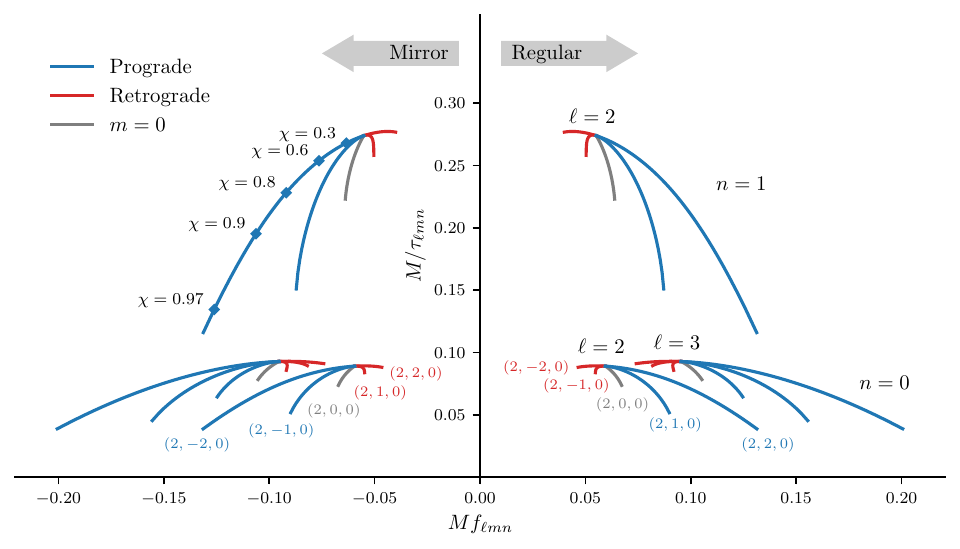}
	\caption{ 
		Selected modes of the Kerr QNM spectrum. BH QNM frequencies are conventionally represented as complex numbers, with the real part giving the angular frequency of the mode and the imaginary part giving (minus) the inverse of the damping time: $\omega_{\ell m n} = 2\pi f_{\ell m n} - i/\tau_{\ell m n}$. Here we plot $f_{\ell m n}$ and $1/\tau_{\ell m n}$, each scaled by the remnant BH mass $M$ to make a dimensionless quantity. The spectrum of a Kerr BH also depends on the dimensionless BH spin magnitude, $\chi$; as the spin is increased from zero, branches with different $m$ grow from the points of the Schwarzschild spectrum.
	}
	\label{fig:qnm_taxonomy}
\end{figure}

A Kerr BH of mass $M$ and dimensionless spin $\chi$ has an infinite number of QNM frequencies $\omega^\pm_{\ell m n} = 2\pi f^\pm_{\ell m n} - i/\tau^\pm_{\ell m n}$, where $\ell$ and $m$ are the usual angular indices and $n$ is an overtone number. 
The $\pm$ superscript indicates the sign of the real part of the frequency, such that $\omega_{\ell m n}^+$ are the ``regular'' QNMs that live in the right half-plane, and $\omega_{\ell m n}^-$ are the ``mirror'' (also referred to as ``twin'' or ``conjugate'') QNMs that live in the left half-plane.
As can be seen in Fig.~\ref{fig:qnm_taxonomy}, there is a symmetry between the
regular and mirror modes given by $\omega^-_{\ell m n} = -\omega^{+*}_{\ell, -m, n}$.
Note that this is distinct from the prograde/retrograde classification also seen in the literature, for which prograde modes satisfy $\mathrm{sgn}(\mathrm{Re}[\omega^\mathrm{prograde}_{\ell m n}]) = \mathrm{sgn}(m)$ and retrograde modes satisfy $\mathrm{sgn}(\mathrm{Re}[\omega^\mathrm{retrograde}_{\ell m n}]) = -\mathrm{sgn}(m)$. 
This classification has the advantage of having a clear physical interpretation; prograde (retrograde) modes are those that rotate in the same (opposite) direction as the remnant BH.
However, this labeling breaks down for the $m=0$ modes and so to avoid any ambiguity we prefer the regular/mirror classification in this work (and in the codes).

We obtain the QNM frequencies in \texttt{qnmfits} by utilizing the open-source Python package \texttt{qnm} which uses a Leaver solver for the radial sector and a spectral eigenvalue technique for the angular sector~\cite{Leaver:1985ax,Cook:2014cta,Stein:2019mop}. 
This approach breaks down for the special $(2,2,8)$ mode, and for this mode, we use data available at Ref.~\cite{GW_Ringdown} (see also Ref.~\cite{Forteza:2021wfq}).
In \texttt{KerrRingdown}, the QNM frequency and spherical-spheroidal mixing coefficients are imported from publicly accessible data available at Ref.~\cite{cook_qnms_url}.

\subsection{Ringdown Waveform}

The ringdown waveform can be written as a sum of spin-weight $-2$ spheroidal-harmonic modes,
\begin{equation} \label{eq:h-QNM-spheroidal}
    h(t,\theta,\phi)=\sum_{\ell m n \pm} \mathcal{A}^{\pm}_{\ell m n} 
    e^{-i\omega^{\pm}_{\ell m n}\left(t-t_{0}\right)} {
    }_{-2} S_{\ell m}(\theta,\phi;a\omega^{\pm}_{\ell m n}),
\end{equation}
where $\mathcal{A}^{\pm}_{\ell m n} = A^\pm_{\ell m n}e^{i\varphi^\pm_{\ell m n}}$ are complex amplitudes and ${}_{-2}S_{\ell m}(\theta,\phi; c)$ are the spheroidal harmonics. 
These take as arguments the polar and azimuthal angles, $\theta$ and $\phi$, and the oblateness parameter $c = a\omega^{\pm}_{\ell m n}$, where $a = M\chi$ is the BH spin.
Note that this expansion is only valid in the rest-frame of the remnant BH, with its spin vector pointing along the positive $z$-axis.
Throughout this discussion, we will assume to be in this frame.
NR is instead usually decomposed into spin-weight $-2$ spherical harmonics,
\begin{equation} \label{eq:h-NR-spherical}
	h(t,\theta,\phi)=\sum_{\ell m} h_{\ell m}(t) {}_{-2}Y_{\ell m}(\theta,\phi),
\end{equation}
providing as data the spherical-harmonic modes $h_{\ell m}$.
To relate the two decompositions, we can write the spheroidal harmonics as an expansion over spin-weight $-2$ spherical harmonics,
\begin{equation} \label{eq:spheroidal-spherical}
	{}_{-2}S_{\ell m}(\theta,\phi; c) = \sum_{\ell'} 
	C_{\ell' \ell m}(c) \ {}_{-2}Y_{\ell' m}(\theta,\phi)
	\,,
  \end{equation}
where $C_{\ell' \ell m}(c)$ are the spherical-spheroidal mixing coefficients ~\cite{Berti:2014fga,Cook:2014cta}. 
Substituting Eq.~\ref{eq:spheroidal-spherical} into Eq.~\ref{eq:h-QNM-spheroidal} and comparing coefficients of the spherical harmonics, we arrive at an expression for the spherical-harmonic modes in terms of QNMs:
\begin{equation}\label{eq:mode-decomp}
	h_{\ell m}(t)=\sum_{\ell' n \pm}\mathcal{A}^{\pm}_{\ell' m n} e^{-i\omega^{\pm}_{\ell' m n}\left(t-t_{0}\right)}C_{\ell \ell' m}(a\omega^{\pm}_{\ell' m n}).
	\end{equation}
This is the model implemented in the codes, and it reveals that each spherical-harmonic mode has contributions from every QNM of the same $m$ (an effect known as mode mixing).
These contributions are weighted by the spherical-spheroidal mixing coefficients, which are also obtained via the \texttt{qnm} package for \texttt{qnmfits} and the data available at Ref.~\cite{cook_qnms_url} for \texttt{KerrRingdown}.

\subsection{Fitting Criteria}

When fitting for the complex amplitudes $\mathcal{A}^{\pm}_{\ell m n}$ the codes aim to minimize the sum of the squares of the residuals between the model and data. That is, a least-squares fit is performed.
It has been shown~\cite{Cook:2020otn} that this is equivalent to minimizing the mismatch
\begin{equation}\label{eq:mismatch}
    \mathcal{M} = 1 -  \frac{|\sum_{\ell m} \langle a_{\ell m} | b_{\ell m} \rangle|} {\sqrt{\left[\sum_{\ell m} \langle a_{\ell m} | a_{\ell m} \rangle \right] \left[\sum_{\ell m} \langle b_{\ell m} | b_{\ell m} \rangle \right]}} ,
\end{equation}
where $a_{\ell m}$ and $b_{\ell m}$ are the spherical-harmonic modes of the model (Eq.~\ref{eq:mode-decomp}) and data (Eq.~\ref{eq:h-NR-spherical}), and we use the following inner product
\begin{equation}\label{eq:inner-product}
    \langle a_{\ell m} | b_{\ell m} \rangle = \int_{t_0}^{t_f} \mathrm{d}t\, a^*_{\ell m}(t) b_{\ell m}(t).
\end{equation}
Internally, the \texttt{qnmfits} package is calling \texttt{numpy.linalg.lstsq}~\cite{Harris:2020xlr} to fit for the complex amplitudes, and \texttt{KerrRingdown} has a custom implementation in terms of normal equations following Ref.~\cite{Cook:2020otn}.
In \texttt{KerrRingdown}, the fitting results can also be obtained by minimizing the mismatch directly, which is referred to as the ``mode-limited eigenvalue method" in the same reference.
If in addition to the amplitudes the remnant mass and spin want to be fitted for, \texttt{qnmfits} performs a minimization of the mismatch via the \texttt{scipy} implementation of the Nelder-Mead algorithm~\cite{NelderMead,Virtanen:2019joe}.
\texttt{KerrRingdown} minimizes the mismatch by performing a coarse search over a grid of mass and spin values, followed by a fine search using Mathematica's \texttt{FindMaximum} function.
Moreover, \texttt{qnmfits} also contains a ``greedy-fit'' algorithm that fits a desired 
number of loudest modes for a specific binary BH system, which helps address the question of 
which modes are important to model in any specific case.

\section{Demonstration}\label{Demonstration}

To demonstrate the codes we perform fits to two NR simulations from the SXS catalog~\cite{Boyle:2019kee}.
These are the GW150914-like \texttt{SXS:BBH:0305}~\cite{SXSCatalog,Lovelace:2016uwp} and \texttt{SXS:BBH\_ExtCCE:0013} (also known as \texttt{q4\_precessing})~\cite{ExtCCECatalog,MaganaZertuche:2021syq} waveforms, produced using SXS's Spectral Einstein Code (SpEC)~\cite{SpECCode}.
Following Ref.~\cite{MaganaZertuche:2021syq}, these waveforms are mapped to the superrest frame~\cite{Moreschi:1988pc,Moreschi:1998mw,Dain:2000lij,Mitman:2021xkq}.

For each waveform we perform three fits (which we label sets 1, 2, and 3) with different QNM and spherical-harmonic content.
The first set is designed to focus on mode mixing; to do this we focus on the $m = 2$ family of modes, fitting the $(\ell, 2, 0, \pm)$ QNMs to the $(\ell, 2)$ spherical-harmonic modes (with $2 \leq \ell \leq 4$).
The second set focuses on overtones by fitting the $(2, 2, n, +)$ QNMs (with $0 \leq n \leq 7$) to the $(2, 2)$ spherical-harmonic mode.
Finally, the third set fits the 20 most important QNMs (as determined by a greedy algorithm~\cite{MaganaZertuche:2021syq}) to all available spherical-harmonic modes with indices $2 \leq \ell \leq 4$ and $-\ell \leq m \leq \ell $. 
The greedy algorithm ranks QNMs according to their power, and we choose to run the algorithm at the time of peak strain.
Note that the regular- and mirror-mode pair ($\pm$) count together as one of the 20 included QNMs.
For \texttt{SXS:BBH:0305}, the chosen modes are
\begin{center}
	\setlength{\tabcolsep}{0.5em} 
	\begin{tabular}{ l l l l l }
		$(2,0,0,\pm)$ & $(2,\pm 1,0,\pm)$ & $(2,\pm 2,0,\pm)$ & $(3,\pm 3,0,\pm)$ & $(4,\pm 4,0,\pm)$ \\[5pt]
		$(2,0,1,\pm)$ & & $(2,\pm 2,1,\pm)$ & $(3,\pm 3,1,\pm)$ & $(4,\pm 4,1,\pm)$ \\[5pt]  
		& & $(2,\pm 2,2,\pm)$ & & $(4,\pm 4,2,\pm)$. \\[5pt] 
	\end{tabular}
\end{center}
For \texttt{SXS:BBH\_ExtCCE:0013}, the chosen modes are
\begin{center}
	\setlength{\tabcolsep}{0.5em} 
	\begin{tabular}{ l l l l l l }
		$(2,0,0,\pm)$ & $(2,\pm 1,0,\pm)$ & $(3,-1,0,\pm)$ & $(2,\pm 2,0,\pm)$ & $(3,\pm 2,0,\pm)$ & $(3,\pm 3,0,\pm)$ \\[5pt]
		$(2,0,1,\pm)$ & $(2,\pm 1,1,\pm)$ & & $(2,\pm 2,1,\pm)$ & $(3,\pm 2,1,\pm)$ & $(3,\pm 3,1,\pm)$ \\[5pt]
		&  & & $(2,2,2,\pm)$ & & \\[5pt]
	\end{tabular}
\end{center}
This demonstrates how the different binary configurations excite different QNMs in the ringdown.

\begin{figure}[t]
	\centering
	\includegraphics[width=\columnwidth]{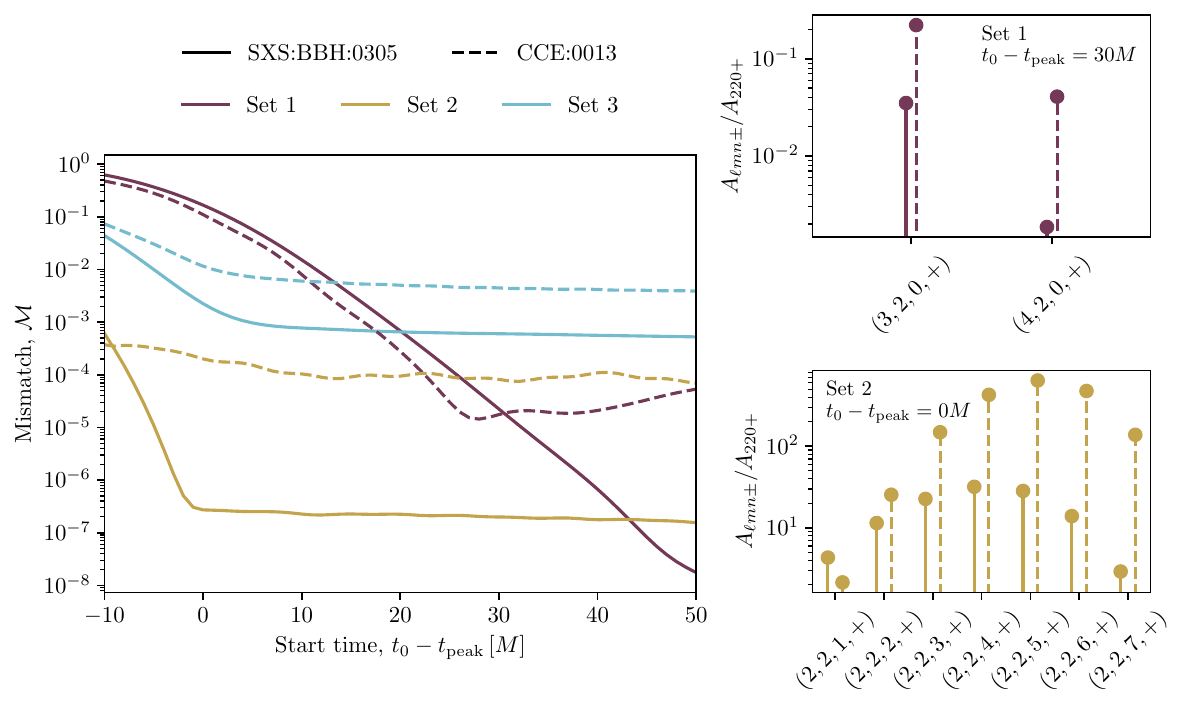}
	\caption{ 
		\textit{Left panel:} Mismatch as a function of the ringdown start time for both simulations considered in this work, and for each set. \textit{Right panels:} QNM amplitudes (normalized by the fundamental mode) at a chosen ringdown start time for both simulations considered in this work. The top panel shows the positive-frequency (prograde) modes from set 1. The bottom panel shows the overtone amplitudes considered in set 2.
	}
	\label{fig:mismatch_amps}
\end{figure}

In the left panel of Fig.~\ref{fig:mismatch_amps} we show the mismatch as a function of the ringdown start time for each set and for both simulations considered. 
When computing the mismatch, the sum in Eq.~\ref{eq:mismatch} is taken to be only over the spherical-harmonic modes included in the fit.
The mismatch is a common metric to measure the quality of fit in the ringdown, and the codes allow it to be easily computed for any combination of data, start time, and QNM content.

The right panels of Fig.~\ref{fig:mismatch_amps} show the magnitude of the QNM amplitudes for Sets 1 (top panel) and 2 (bottom panel), normalized by the fundamental mode amplitude, at a chosen ringdown start time.
As in the mismatch curve, solid lines are for the simulation \texttt{SXS:BBH:0305} and dashed lines are for \texttt{SXS:BBH\_ExtCCE:0013} (and it can clearly be seen that the precessing \texttt{SXS:BBH\_ExtCCE:0013} excited the higher order $(4,2,0,+)$ mode more than the aligned-spin \texttt{SXS:BBH:0305}, as expected).
QNM amplitudes are an ongoing area of study, and via least-squares fitting the codes give (essentially instantaneous) estimates of the amplitudes.

As part of this work, it has been verified that the Python and Mathematica packages produce identical results, up to machine precision, for both the mismatch and complex QNM amplitudes for a variety of test cases.

\section{Conclusions}\label{Conclusions}


Understanding the relationship between QNM amplitude excitations and binary BH 
configurations provides a good way to test GR in the strong field regime. Fitting for these
amplitudes, however, can quickly become a difficult task. To alleviate some of these difficulties 
we introduce \texttt{qnmfits} and \texttt{KerrRingdown}, which allow for the extraction of the 
complex QNM amplitudes for any waveform data provided. These codes (in Python and Mathematica, 
respectively) allow the user to easily specify the time of QNM amplitude extraction, as well as 
which QNMs and spherical harmonics to take into account. Both codes also have the capability to fit for the remnant 
mass and spin of the BH. 

It should be noted that there are known limitations to the least-squares fitting approach currently adopted by the codes. 
In particular, the core fitting functions in the codes do not come equipped with a notion of fitting uncertainty, simply returning a single `best fit' solution for the QNM amplitudes.
For relatively loud and long-lived QNMs this is usually not a problem, but care should be taken when fitting for rapidly decaying or subdominant QNMs; results can be biassed by numerical error in the waveform being fitted to~\cite{Clarke:2024lwi,Ferguson:2020xnm,Carullo:2023tff,Jan:2023raq,Wang:2024iyj}, or unmodelled features in the waveform such as tails~\cite{Carullo:2023tff} or subdominant QNMs.
Nevertheless, least-squares fits are an important tool due to their speed and ease of use, and we hope the codes presented here can aid in future studies of the ringdown.

%

\section*{Acknowledgments}

L.M.Z., L.G., and E.F. all contributed equally to this work.
The authors would like to thank Leo Stein and Chris Moore for their guidance and 
fruitful discussions. They also thank Keefe Mitman for clarifications on supertranslations. 
Some calculations were performed with the Wheeler cluster at the California Institute of Technology 
(Caltech), which is supported by the Sherman Fairchild Foundation and by Caltech.
Some computations using the \texttt{KerrRingdown} package were performed on the Wake Forest University DEAC Cluster~\cite{DEAC-Cluster}, a centrally managed resource with support provided in part by the University.
The work of L.M.Z. was partially supported by the MSSGC Graduate
Research Fellowship, awarded through the NASA Cooperative Agreement
80NSSC20M0101.
All plots were made using the Python package \texttt{matplotlib}~\cite{Hunter:2007ouj}.

\section*{References}
\bibliographystyle{iopart-num}
\bibliography{qnm_comparison}

\providecommand{\newblock}{}
\begin{thebibliography}{10}
\expandafter\ifx\csname url\endcsname\relax
  \def\url#1{{\tt #1}}\fi
\expandafter\ifx\csname urlprefix\endcsname\relax\def\urlprefix{URL }\fi
\providecommand{\eprint}[2][]{\url{#2}}

\bibitem{Teukolsky:1972my}
Teukolsky S~A 1972 {\em Phys. Rev. Lett.\/} {\bf 29} 1114--1118

\bibitem{Teukolsky:1973}
Teukolsky S~A 1973 {\em Astrophys. J.\/} {\bf 185} 635--648

\bibitem{Press:1973zz}
Press W~H and Teukolsky S~A 1973 {\em Astrophys. J.\/} {\bf 185} 649--674

\bibitem{Finch:2023}
Finch E 2023 {\em {Black-hole Ringdown: Quasinormal Modes in
  Numerical-relativity Simulations and Gravitational-wave Observations}\/}
  Ph.D. thesis \urlprefix\url{https://etheses.bham.ac.uk/id/eprint/13992}

\bibitem{Giesler:2019uxc}
Giesler M, Isi M, Scheel M~A and Teukolsky S 2019 {\em Phys. Rev. X\/} {\bf 9}
  041060 (\textit{Preprint} \eprint{1903.08284})

\bibitem{Dhani:2020nik}
Dhani A 2021 {\em Phys. Rev. D\/} {\bf 103} 104048 (\textit{Preprint}
  \eprint{2010.08602})

\bibitem{Cheung:2021bol}
Cheung M~H~Y, Destounis K, Macedo R~P, Berti E and Cardoso V 2022 {\em Phys.
  Rev. Lett.\/} {\bf 128} 111103 (\textit{Preprint} \eprint{2111.05415})

\bibitem{Berti:2022xfj}
Berti E, Cardoso V, Cheung M~H~Y, Di~Filippo F, Duque F, Martens P and
  Mukohyama S 2022 {\em Phys. Rev. D\/} {\bf 106} 084011 (\textit{Preprint}
  \eprint{2205.08547})

\bibitem{Cardoso:2024mrw}
Cardoso V, Kastha S and Panosso~Macedo R 2024  (\textit{Preprint}
  \eprint{2404.01374})

\bibitem{Redondo-Yuste:2023ipg}
Redondo-Yuste J, Pere\~niguez D and Cardoso V 2024 {\em Phys. Rev. D\/} {\bf
  109} 044048 (\textit{Preprint} \eprint{2312.04633})

\bibitem{Zhu:2024dyl}
Zhu H {\em et~al.\/} 2024  (\textit{Preprint} \eprint{2404.12424})

\bibitem{Mitman:2022qdl}
Mitman K {\em et~al.\/} 2023 {\em Phys. Rev. Lett.\/} {\bf 130} 081402
  (\textit{Preprint} \eprint{2208.07380})

\bibitem{Cheung:2022rbm}
Cheung M~H~Y {\em et~al.\/} 2023 {\em Phys. Rev. Lett.\/} {\bf 130} 081401
  (\textit{Preprint} \eprint{2208.07374})

\bibitem{London:2014cma}
London L, Shoemaker D and Healy J 2014 {\em Phys. Rev. D\/} {\bf 90} 124032
  [Erratum: Phys.Rev.D 94, 069902 (2016)] (\textit{Preprint}
  \eprint{1404.3197})

\bibitem{London:2018gaq}
London L~T 2020 {\em Phys. Rev. D\/} {\bf 102} 084052 (\textit{Preprint}
  \eprint{1801.08208})

\bibitem{Cheung:2023vki}
Cheung M~H~Y, Berti E, Baibhav V and Cotesta R 2024 {\em Phys. Rev. D\/} {\bf
  109} 044069 (\textit{Preprint} \eprint{2310.04489})

\bibitem{Hamilton:2023znn}
Hamilton E, London L and Hannam M 2023 {\em Phys. Rev. D\/} {\bf 107} 104035
  (\textit{Preprint} \eprint{2301.06558})

\bibitem{Zhu:2023fnf}
Zhu H {\em et~al.\/} 2023  (\textit{Preprint} \eprint{2312.08588})

\bibitem{OShaughnessy:2012iol}
O'Shaughnessy R, London L, Healy J and Shoemaker D 2013 {\em Phys. Rev. D\/}
  {\bf 87} 044038 (\textit{Preprint} \eprint{1209.3712})

\bibitem{Pratten:2020ceb}
Pratten G {\em et~al.\/} 2021 {\em Phys. Rev. D\/} {\bf 103} 104056
  (\textit{Preprint} \eprint{2004.06503})

\bibitem{Baibhav:2023clw}
Baibhav V, Cheung M~H~Y, Berti E, Cardoso V, Carullo G, Cotesta R, Del~Pozzo W
  and Duque F 2023  (\textit{Preprint} \eprint{2302.03050})

\bibitem{Ma:2022wpv}
Ma S, Mitman K, Sun L, Deppe N, H\'ebert F, Kidder L~E, Moxon J, Throwe W, Vu
  N~L and Chen Y 2022 {\em Phys. Rev. D\/} {\bf 106} 084036 (\textit{Preprint}
  \eprint{2207.10870})

\bibitem{Redondo-Yuste:2023seq}
Redondo-Yuste J, Carullo G, Ripley J~L, Berti E and Cardoso V 2024 {\em Phys.
  Rev. D\/} {\bf 109} L101503 (\textit{Preprint} \eprint{2308.14796})

\bibitem{Carullo:2023tff}
Carullo G and De~Amicis M 2023  (\textit{Preprint} \eprint{2310.12968})

\bibitem{Clarke:2024lwi}
Clarke T~A {\em et~al.\/} 2024  (\textit{Preprint} \eprint{2402.02819})

\bibitem{qnmfitscode}
Magaña~Zertuche L and Finch E 2025 qnmfits
  \urlprefix\url{https://doi.org/10.5281/zenodo.14806974}

\bibitem{KerrRingdownCode}
Cook G~B and Gao L 2025 Kerr{R}ingdown
  \urlprefix\url{https://doi.org/10.5281/zenodo.14804284}

\bibitem{Boyle:2019kee}
Boyle M {\em et~al.\/} 2019 {\em Class. Quant. Grav.\/} {\bf 36} 195006
  (\textit{Preprint} \eprint{1904.04831})

\bibitem{Healy:2022wdn}
Healy J and Lousto C~O 2022 {\em Phys. Rev. D\/} {\bf 105} 124010
  (\textit{Preprint} \eprint{2202.00018})

\bibitem{Hamilton:2023qkv}
Hamilton E {\em et~al.\/} 2024 {\em Phys. Rev. D\/} {\bf 109} 044032
  (\textit{Preprint} \eprint{2303.05419})

\bibitem{Ferguson:2023vta}
Ferguson D {\em et~al.\/} 2023  (\textit{Preprint} \eprint{2309.00262})

\bibitem{Boyle_The_sxs_package_2025}
Boyle M and Scheel M 2025 The sxs package
  \urlprefix\url{https://doi.org/10.5281/zenodo.14776832}

\bibitem{scri_url}
Boyle M, Iozzo D and Stein L~C 2020 moble/scri: v1.2
  \urlprefix\url{https://doi.org/10.5281/zenodo.4041972}

\bibitem{MaganaZertuche:2021syq}
Maga\~na Zertuche L {\em et~al.\/} 2022 {\em Phys. Rev. D\/} {\bf 105} 104015
  (\textit{Preprint} \eprint{2110.15922})

\bibitem{Leaver:1985ax}
Leaver E~W 1985 {\em Proc. Roy. Soc. Lond. A\/} {\bf 402} 285--298

\bibitem{Cook:2014cta}
Cook G~B and Zalutskiy M 2014 {\em Phys. Rev. D\/} {\bf 90} 124021
  (\textit{Preprint} \eprint{1410.7698})

\bibitem{Stein:2019mop}
Stein L~C 2019 {\em J. Open Source Softw.\/} {\bf 4} 1683 (\textit{Preprint}
  \eprint{1908.10377})

\bibitem{GW_Ringdown}
Forteza X~J and Mourier P 2024 accessed: 2024-09-18
  \urlprefix\url{https://codeberg.org/GW_Ringdown}

\bibitem{Forteza:2021wfq}
Forteza X~J and Mourier P 2021 {\em Phys. Rev. D\/} {\bf 104}(12) 124072
  \urlprefix\url{https://link.aps.org/doi/10.1103/PhysRevD.104.124072}

\bibitem{cook_qnms_url}
Cook G~B 2024 Kerr modes: Phase fixed gravitational qnms and ttms
  \urlprefix\url{https://doi.org/10.5281/zenodo.2650357}

\bibitem{Berti:2014fga}
Berti E and Klein A 2014 {\em Phys. Rev. D\/} {\bf 90} 064012
  (\textit{Preprint} \eprint{1408.1860})

\bibitem{Cook:2020otn}
Cook G~B 2020 {\em Phys. Rev. D\/} {\bf 102} 024027 (\textit{Preprint}
  \eprint{2004.08347})

\bibitem{Harris:2020xlr}
Harris C~R {\em et~al.\/} 2020 {\em Nature\/} {\bf 585} 357--362
  (\textit{Preprint} \eprint{2006.10256})

\bibitem{NelderMead}
Gao F and Han L 2012 {\em Computational Optimization and Applications\/} {\bf
  51} 259--277 \urlprefix\url{https://doi.org/10.1007/s10589-010-9329-3}

\bibitem{Virtanen:2019joe}
Virtanen P {\em et~al.\/} 2020 {\em Nature Meth.\/} {\bf 17} 261
  (\textit{Preprint} \eprint{1907.10121})

\bibitem{SXSCatalog}
{SXS Gravitational Waveform Database}
  \url{http://www.black-holes.org/waveforms}

\bibitem{Lovelace:2016uwp}
Lovelace G {\em et~al.\/} 2016 {\em Class. Quant. Grav.\/} {\bf 33} 244002
  (\textit{Preprint} \eprint{1607.05377})

\bibitem{ExtCCECatalog}
{SXS Ext-CCE Waveform Database}
  \url{https://data.black-holes.org/waveforms/extcce_catalog.html}

\bibitem{SpECCode}
\url{https://www.black-holes.org/code/SpEC.html}

\bibitem{Moreschi:1988pc}
Moreschi O~M 1988 {\em Class. Quant. Grav.\/} {\bf 5} 423--435

\bibitem{Moreschi:1998mw}
Moreschi O~M and Dain S 1998 {\em J. Math. Phys.\/} {\bf 39} 6631--6650
  (\textit{Preprint} \eprint{gr-qc/0203075})

\bibitem{Dain:2000lij}
Dain S and Moreschi O~M 2000 {\em Class. Quant. Grav.\/} {\bf 17} 3663--3672
  (\textit{Preprint} \eprint{gr-qc/0203048})

\bibitem{Mitman:2021xkq}
Mitman K {\em et~al.\/} 2021 {\em Phys. Rev. D\/} {\bf 104} 024051
  (\textit{Preprint} \eprint{2105.02300})

\bibitem{Ferguson:2020xnm}
Ferguson D, Jani K, Laguna P and Shoemaker D 2021 {\em Phys. Rev. D\/} {\bf
  104} 044037 (\textit{Preprint} \eprint{2006.04272})

\bibitem{Jan:2023raq}
Jan A, Ferguson D, Lange J, Shoemaker D and Zimmerman A 2024 {\em Phys. Rev.
  D\/} {\bf 110} 024023 (\textit{Preprint} \eprint{2312.10241})

\bibitem{Wang:2024iyj}
Wang Z, Zhao J and Cao Z 2024 {\em Commun. Theor. Phys.\/} {\bf 76} 015403
  (\textit{Preprint} \eprint{2401.15331})

\bibitem{DEAC-Cluster}
 2021 {WFU High Performance Computing Facility}
  \urlprefix\url{https://hpc.wfu.edu}

\bibitem{Hunter:2007ouj}
Hunter J~D 2007 {\em Comput. Sci. Eng.\/} {\bf 9} 90--95

\end{thebibliography}

\end{document}